# A Promising Ohmic Contacts Approach for High-Al $Al_xGa_{1-x}N$ (x>0.6) Channel HEMTs with AlN/GaN Digital Alloy Channel

Tariq Jamil[1,a)], Abdullah Al Mamun Mazumder[1,a)], S M Tazbiul Hasan[1], Mafruda Rahman[1], Muhammad Ali[1], Ankit Malik[1], Kamal Hussain[2], Chandan Joishi[3], Mansura Sadek[3], James G. Fiorenza[3], Grigory Simin[1], and Asif Khan[1]

[1]Department of Electrical Engineering, University of South Carolina, Columbia, SC 29208, USA

[2]Currenttly @ Texas Instruments Incorporated, Dallas, TX 75243, USA

[3]Analog Devices Incorporated, Boston, MA 02110, USA

[a)] **Contributions:** T. Jamil and A.A.M. Mazumder contributed equally to this work.

**Authors to whom correspondence should be addressed:** tjamil@email.sc.edu, mazumdea@email.sc.edu

In this paper we report a novel ohmic contact formation scheme for Extreme Bandgap (EBG) $Al_xGa_{1-x}N$ (x>0.6) channel HEMTs with undoped barrier layers. Our approach consists of using a new low temperature (LT) pulsed metal-organic chemical vapor deposition (PMOCVD) doping scheme for the $n^{++}$-GaN regrown contacts and an $Al_xGa_{1-x}N$ digital alloy (DA) channel layer comprising short period super lattices (SPSL) of AlN and GaN. Pulsed growth and doping yield a sheet resistivity which is a factor of 3-5 lower than that of conventional doped $n^{++}$-GaN layers grown under identical conditions. Moreover, the regrown $n^{++}$-GaN layer has no hetero-barrier with the GaN layers of the $Al_xGa_{1-x}N$ DA channel. These innovations led to MOCVD regrown linear ohmic contacts and a record-low contact resistance $R_c$ ~6.5 Ω-mm to the $Al_{0.62}Ga_{0.38}N$ DA channel layer of a HEMT with AlN barrier layer.

Extreme Bandgap (EBG) $Al_xGa_{1-x}N$ ($0.7 < x < 1$ and a bandgap ~ 5 to 6.2 eV) semiconductors have a high breakdown field of 8 to 15 MV/cm, making them an attractive choice for high-power electronic devices in extreme environments. Due to their high channel currents and power-handling capability in a compact footprint, which is enabled by the high channel conductivity and the large critical breakdown electric field,[1-4] Ultra-wide bandgap (UWBG) and EBG high electron mobility transistors (HEMTs) have been explored for various high-power electronic applications.[5-7] Several groups, including ours, have reported EBG AlGaN channel HEMTs with the channel alloy compositions of x ≈ (0.60 - 0.75) on either AlN/sapphire templates or bulk AlN substrates.[8-17] In these reported structures, the channel and barrier layers are random $Al_xGa_{1-x}N$ alloys, and the barrier layer composition 'x' is typically ~ 0.85. Nearly for all the reported cases, the barrier layers were n-type doped to assist with the formation of linear ohmic contacts.[8, 11, 18] However, the high n-type barrier doping invariably leads to a reduction in the device breakdown field.[10] In addition to doping, the linearity of the ohmic contact and its resistance are also impacted by the barrier layer composition. Making direct metallic contact to even a highly doped barrier layer with alloy composition x ~ 0.85 is challenging due to the large Schottky barrier height resulting from the higher alloy composition.

To avoid this, nearly all the research groups working on EBG AlGaN channel HEMTs employ reverse composition graded $Al_xGa_{1-x}N$ layers (RGL), which are highly silicon-doped to overcompensate for the polarization doping, leading to the generation of holes in the RGL.[13, 14] Using this RGL surface contact formation approach, Mamun et al. reported contact resistance ($R_c$) of ~ 4.3 Ω-mm for devices with $Al_{0.65}Ga_{0.35}N$ channel and $Al_{0.85}Ga_{0.15}N$ barrier layers.[8] Recently, Gohel et al. reported $Al_{0.65}Ga_{0.35}N$-channel HEMTs using a Si-doped $Al_{0.85}Ga_{0.15}N$ barrier and a 50-nm-thick reverse graded $Al_xGa_{1-x}N$ (x~ 0.84 to 0.40) contact layer, achieving a substantially reduced contact resistance of ~ 2.6 Ω-mm.[11] Similarly, Zhu/Shin et al. achieved a low contact resistance of 1.55 Ω-mm (contact resistivity $1.4 \times 10^{-6}$ Ω-cm$^2$), using RGL heterostructure and interfacial engineering approaches.[14, 19] For all these reports, the doped barrier design with n-doped RGL not only leads to a reduction in the breakdown field but also requires a precise etching of the RGL layer from the access region. Over-etching reduces the peak current, and under-etching leads to excessive gate leakage current. The use of doped barrier layers also reduces the channel mobility, as the incorporation of Si dopants in the barrier adds significant ionized-impurity scattering.[20, 21] Mukhopadhyay et al. showed that in $Al_{0.65}Ga_{0.35}N$-channel HEMTs, the 2DEG mobility is primarily limited by alloy-disorder scattering and ionized-impurity scattering from Si-doped barriers, even when low contact resistances and high breakdown voltages are achieved.[18]

One potential approach to avoid barrier doping and the use of contact formation n-doped RGL layers is to use selective area regrown $n^{++}$-GaN contact layers. These are commonly used for low-resistance ohmic contacts to GaN-channel HEMTs. Molecular beam epitaxy (MBE) and MOCVD-regrown $n^{++}$-GaN contacts have been demonstrated for GaN-channel HEMTs.[22-31)] However, to date, there are a very limited number of reports of selective area regrown $n^{++}$-GaN contacts for UWBG and EBG AlGaN channel HEMTs using either MBE or MOCVD. Singhal et al. reported $AlN/Al_xGa_{1-x}N/AlN$ HEMTs (x = 0 .25, 0.44, and 0.58), utilizing MBE $n^{++}$-GaN regrowth, with $R_c$ ~ 0.22, 9.3, 212.1 Ω-mm, respectively.[32] Also using MBE regrowth, Abid et al. reported $AlN/Al_{0.5}Ga_{0.5}N/AlN$ HEMTs with selectively regrown $n^{++}$-GaN Ohmic contacts, achieving reduced contact resistances of ~21-58 Ω-mm.[33] With MOCVD $n^{++}$-GaN regrowth, Baca et al. reported $AlN/Al_{0.85}Ga_{0.15}N$ HEMTs for which the source/drain contacts remained strongly nonlinear, with an estimated contact resistance on the order of ~1900 Ω·mm.[34] Also with MOCVD regrowth, Douglas et al. reported $Al_{0.85}Ga_{0.15}N/Al_{0.66}Ga_{0.34}N$ HEMTs with an $R_c$ ~ 62 Ω-mm.[35] Kometani et al. demonstrated $AlN/Al_{0.72}Ga_{0.28}N$ HEMTs on single-crystal AlN by MOCVD with direct contact and an extracted contact resistance of ~ 200 Ω-mm, highlighting the difficulty of ohmic contact formation for high-Al content $Al_xGa_{1-x}N$.[36] As seen from these references, no group to date has been able to form low-contact-resistance linear ohmic contacts to the EBG $Al_xGa_{1-x}N$ channel HEMTs using regrown $n^{++}$-GaN layers grown with either MOCVD or MBE. Maeda et al. have reported $AlN/Al_{0.5}Ga_{0.5}N$ HEMTs with regrown $n^{++}$-GaN contacts with $R_c$ ~ 0.43 Ω-mm.[37] However, they used LT pulsed sputter deposition for the regrowth, and their channel composition was only x ~ 0.5.

In this paper, we report a novel ohmic contact formation scheme for EBG $Al_xGa_{1-x}N$ ($x > 0.6$) channel HEMTs with undoped AlN barrier layers. Our approach consists of using a new low-temperature (LT) pulsed MOCVD doping scheme for the $n^{++}$-GaN selective area regrown contacts and $Al_xGa_{1-x}N$ DA channel layers comprising short-period super lattices (SPSL) of AlN and GaN. The LT pulsed MOCVD regrown $n^{++}$-GaN layer was measured to have an electron concentration and mobility of $1.2 \times 10^{20}$ $cm^{-3}$, and ~ 30 $cm^2$ $V^{-1}s^{-1}$, respectively, which translates to a sheet resistance that is a factor of 3-5 lower than that for conventional MOCVD regrowth under identical growth conditions. Moreover, the regrown $n^{++}$-GaN layer has no hetero-barrier with the GaN layers of the $Al_xGa_{1-x}N$ DA channel. These innovations led to a record low contact resistance, $R_c$ ~ 6.5 Ω-mm for a linear ohmic contact to the $Al_{0.62}Ga_{0.38}N$ DA channel layer of a HEMT with an undoped AlN barrier layer.

The AlN/GaN SPSL-based $Al_xGa_{1-x}N$ digital alloy approach was first demonstrated by Khan/Mcmillan et al.[38] Following this, Nikishin et al. and Sun et al. used AlGaN DA layers for the active region of a deep UV light-emitting diode.[39,40,41] DA $Al_xGa_{1-x}N$ channel comprises thin AlN and GaN layers with thicknesses *t*(AlN) and *t*(GaN) repeated say *R* times (see Fig. 1(a)).

The fundamental advantage of a DA channel over a conventional (random alloy (RA)) channel for HEMT devices is as follows. As shown in [41], if the individual layers' thicknesses in the DA do not exceed a few monolayers (ML), the carrier wave functions in the direction perpendicular to the layer plane tunnel through the barrier regions and couple with one another. This phenomenon creates the minibands of the DA.[41] As a result, in the direction perpendicular to the layer's plane, the DA acts as a single $Al_xGa_{1-x}N$ layer whose thickness and composition are given by:[40, 41]

$$T = [t(AlN) + t(GaN)] \times R \qquad (1)$$

$$(Al\%) = \frac{t(AlN)}{t(AlN) + t(GaN)} \times 100 \qquad (2)$$

This equivalency manifests itself in effective bandgap, optical properties, XRD, and other characteristics. Fig. 1(b) compares the extracted carrier distribution for 62% digital-alloy AlN/GaN HEMTs and 62% random-alloy AlGaN HEMTs by 1D Schrödinger–Poisson solver by G. Shnider[42]), showing a sharper, higher-amplitude peak for the digital-alloy channel.

In the lateral direction, however, carriers in the DA channel are not confined. The electron wave function for each individual GaN layer has a peak within that layer. Therefore, in terms of lateral electrical transport and contact formation, the GaN sublayers of the DA still act mostly as GaN material, except the energy level being shifted due to adjacent AlN barriers. These anisotropic characteristics of DA in lateral and transverse directions are fundamental to our approach to create a high Al-content channel layer and, at the same time, to enable linear ohmic contacts between regrown $n^{++}$-GaN contacts and GaN sub-layers of the DA.

Currently, for most of the $Al_xGa_{1-x}N$ ultrawide bandgap HEMTs, a channel composition around x ~0.65 is employed. This enables the use of a highly doped $n^{++}$-$Al_{0.85}Ga_{0.15}N$ barrier layer. The growth-rate/metalorganic (MO) switching and sweepout time capability of our MOCVD system allows us to grow a monolayer of AlN or GaN with a high confidence level. To decide the thicknesses of the AlN/GaN superlattice unit cell composition of the DA channel layer to yield a channel composition close to 65%, $Al_xGa_{1-x}N$ DA layers with varying AlN and GaN thicknesses (number of monolayers in the unit cell) were grown on high-quality, low-defect AlN (2.6 μm)/sapphire templates. In each case, the unit cell was repeated 100 times. The number of GaN monolayers was selected to be either 1 or 2, and the AlN monolayer number varied from 1 to 3 to 9. Then their room-temperature CL spectra were measured using a 20 keV beam. These CL data are plotted in Fig. 4(c) along with the estimated bandgap energy values. These estimated bandgap values are shown in Fig. 1(d). Also included in Fig. 1(d) is a curve corresponding to the estimated Al-alloy composition for the DA using Eq. 2 above and then converting to the bandgap energy using Vegard's law with a bowing parameter b ~0.7.[43, 44]

$$E_g(x) = (1-x)E_{g,GaN} + xE_{g,AlN} - bx(1-x) \quad (3)$$

As seen from Fig. 1(d), there is good agreement between the measured and estimated alloy compositions. Based on these CL data, we selected a DA SPSL configuration with 2 ML (~ 5 Å) of GaN and 3 ML (~ 7.5 Å AlN) to yield a channel composition close to 65%.

Then an SPSL DA (AlN/GaN) channel HEMT epilayer with an AlN barrier layer was grown on a 2.5 µm thick AlN template on a 2-inch diameter c-plane (0001) sapphire substrate. The details of the AlN template growth and characterization are reported elsewhere.[45] A schematic of the epilayer structure for the DA channel HEMT is shown in Fig. 2(a). As seen, it comprises a 300 nm thick undoped buffer on the AlN template, followed by the 250 nm $Al_{0.67}Ga_{0.33}N$ back barrier layer, followed by a 22 nm thick AlN/GaN SPSL ($Al_{0.64}Ga_{0.36}N$) channel, and a 25 nm thick AlN barrier layer. The targeted thicknesses of the AlN and GaN layers of the channel SPSL were 7.5 Å and 5 Å, leading to an average Al composition of 64%. This selection was made based on the CL data of Fig. 1(c). The high-Al (x ~ 0.67) $Al_xGa_{1-x}N$ back barrier provides strong vertical electron confinement, suppresses buffer leakage and punch-through, and helps to avoid back depletion. The growth temperatures for the AlN buffer and $Al_{0.67}Ga_{0.33}N$ layer were 1100 and 1150 ºC, respectively, while the AlGaN DA channel layer was grown at 1050 ºC. High-resolution X-ray diffraction on-axis (0002) measurements show the Al-alloy composition of the channel and the back barrier layers to be 62% and 67%. This data is included in Fig. 2(b).

The high-resolution cross-section TEM data for the epi-layer structure of the DA channel HEMT is shown in Fig. 3. Also included in Fig. 3 is the zoom cross-section of the DA GaN/AlN SPSL layers. From this data, we confirmed that the total thickness of the SPSL channel is 22 nm, with each unit cell comprising 7 Å of AlN and 5 Å of GaN.

In Fig. 4(a), we have included the surface atomic force microscope (AFM) scans for the DA channel HEMT layers. An RMS surface roughness value of 0.4 nm was measured for a 2 × 2 µm scan. In Fig. 4(b) the sheet resistance mapping of the 2-inch SPSL HEMT is shown. The average sheet resistance was around 2250 Ω/□ with 11% standard deviation. We measured room-temperature (RT) capacitance-voltage (C-V) of the epi-layers using the Hg-probe C-V system (Material Development Corporation), included in Fig. 4(c). From it, the extracted carrier concentration and mobility are $2.7 \times 10^{13}$ $cm^{-2}$, and 105 $cm^{-2}$/V-sec, respectively. For these extractions, we used the following capacitance-voltage relations for the HEMT structure, Eq(4):[46, 47]

$$Q = qN_sA = C \times \mathrm{V_{th}} \qquad (4)$$

$$N_s = \frac{C\mathrm{V_{th}}}{Aq}$$

$$\mu = \frac{1}{qN_s\mathrm{R_{sh}}}$$

Where $Q$ is the total charge associated with the 2DEG, q = 1.602 $\times 10^{-19}$ C is the electronic charge, $V_{th}$ is the threshold voltage, $C$ is the structure capacitance above the threshold voltage, $N_s$ is the sheet electron density in the 2DEG, $A$=3.74 $\times 10^{-3}$ cm$^{-2}$ is the mercury probe area, $\mu$ is electron mobility in the 2DEG, and $R_{sh}$ is the sheet resistance found from contactless sheet resistance mapping system (Lehighton). Furthermore, van der Pauw Hall measurements were also performed; the carrier concentration and mobility are $3 \times 10^{13}$ cm$^{-2}$, and 95 cm$^{-2}$/V-sec, respectively, which show good agreement with the extracted carrier concentration and mobility values from CV. The extracted mobility value of 105 cm$^2$/V-sec is higher than other reports of AlN barrier HEMTs with random $Al_xGa_{1-x}N$ alloy channels. Singhal et al. reported a mobility of 24 cm$^2$/V-sec for an $Al_{0.58}Ga_{0.42}N$ channel, while Komentani et al. reported a mobility of 72 cm$^2$/V-sec for an $Al_{0.72}Ga_{0.28}N$ channel.[32, 36] The achieved higher mobility can be attributed to the absence of alloy-disorder scattering in the DA channel, which consists of GaN and AlN layers. More studies of the carrier field effect mobility will be carried out using HEMT devices with DA channels and will be reported later.

Finally, we fabricated $n^{++}$-GaN regrown ohmic contacts in a TLM pattern to connect to the channel layers. For this, first, a $Si_3N_4$ hard mask was deposited using PECVD. Then we use the standard photolithography and ICP-RIE etching process to open the TLM ohmic pad regions. Then, using a low-temperature (LT) pulsed MOCVD doping scheme, a highly Si-doped $n^{++}$-GaN layer was grown into the trench regions for the TLM pads. In this pulsed growth mode, the Ga-, N-, and Si-precursors, namely trimethylgallium (TMGa), ammonia ($NH_3$), and silane ($SiH_4$), are introduced in the chamber alternately to reduce adduct formation and improve surface morphology. The $n^{++}$-GaN contact layer was grown at 830 ºC, and a chamber pressure of 70 torr. From a separate flat growth on an AlN template, using van der Pauw Hall measurements effects under a magnetic field of 0.3 T (MMR Technologies system), we verified that the pulsed doping procedure yields an $n^{++}$-GaN layer with an electron concentration and mobility of $1.2 \times 10^{20}$ cm$^{-3}$ and 30 cm$^2$ V$^{-1}$s$^{-1}$, respectively. Its sheet resistance is a factor of 3-5 lower than that for conventional MOCVD regrowth with the same $SiH_4$ flow.

After the $n^{++}$-GaN regrowth, we removed the $Si_3N_4$ hard mask using wet chemical etching. Subsequently, Ti/Al/Ti/Au (15 nm/70 nm/30 nm/50 nm) ohmic contacts were deposited using e-beam evaporation in the TLM pattern. The standard rectangular geometry transmission line model (TLM) pattern was 200 µm wide with contact spacings ranging from 8 to 20 µm. All the TLM measurements were carried out using ICP-RIE-isolated MESAs to eliminate current spreading, which would lead to inaccurate sheet and contact resistance measurements. The I-V data for the TLM pattern is included in Fig. 5(a) and an inset to the figure corresponding to the spacing versus resistance. As seen from the data, the I-V curves are completely linear, and the extracted $R_c$ and $R_{sh}$ values were 6.5 Ω mm and 2503 Ω/□. This sheet resistance value is in good agreement with that measured using contactless Lehighton mapping (Fig. 4b). In Fig. 5 (b), we have included TLM data like that of Fig. 5(a) for the case where direct surface contacts were fabricated on the DA channel HEMT epilayers on isolated MESAs. As seen, the direct surface contacts were highly nonlinear with approximately a 6-volt barrier. This is like what was observed in other reports on UWBG channel HEMTs with undoped high-Al composition $Al_xGa_{1-x}N$ barriers.[36, 48, 49] The extracted contact is 33.8 Ω-mm, with high sheet resistance of 5049 Ω/□, measured from the linear part of the I-V curves. The sheet resistance is higher than that of the as-grown sample (~2250 Ω/□). This substantial deviation indicates that direct surface contacts are ineffective for fully contacting the 2-DEG.

In Fig. 5 (c), we have included the benchmark comparison for channel Al-composition versus $R_c$ for other reported MBE and MOCVD regrown contacts with the channel $Al_xGa_{1-x}N$ Al-alloy composition around 60% or higher. This comparison demonstrates that our $R_c$ is the lowest among all reported high Al-composition $Al_xGa_{1-x}N$ channel HEMTs using either MOCVD or MBE regrown $n^{++}$-GaN contacts. In Fig. 5 (d), the $R_c$ values are plotted as a function of the barrier alloy composition. As seen, our results represent the lowest contact resistance when compared to other reported MBE/MOCVD regrown contacts.

In summary, we report a new ohmic contact scheme for EBG $Al_xGa_{1-x}N$ ($x>0.6$) HEMTs with undoped AlN barrier layers by employing an $Al_xGa_{1-x}N$ digital alloy (DA) channel and using pulsed MOCVD Si-doping for the regrown $n^{++}$-GaN layers. These innovations enabled linear ohmic contacts with record low contact resistance of $R_c$ ~ 6.5 Ω-mm. The use of undoped barriers in our regrown contacts approach should in principle enable higher breakdown fields for the HEMT devices.

**Acknowledgment**

This project is funded by the Microelectronics Commons Program, a DoW initiative. This material is based upon work supported by ADI under award number 10015113. The views expressed are those of the author and do not reflect the official policy or position of the Department of War or the U.S. Government. Our work is also supported by the ONR grant number N000142312010. (Capt. Lynn Petersen). The reported work is also partially supported by the Air Force Office of Scientific Research under award number FA9550-24-1-0269. Any opinions, findings, and conclusions or recommendations expressed in this material are those of the author(s) and do not necessarily reflect the views of the United States Air Force (Dr. Ali Sayir). This work is also supported by the UofSC Research Institute for Extreme Electronics, USC grant number 80005646.

## AUTHOR DECLARATIONS

### Conflict of Interest

The authors have no conflicts to disclose.

## DATA AVAILABILITY

The data that support the findings of this study are available from the corresponding author upon reasonable request.

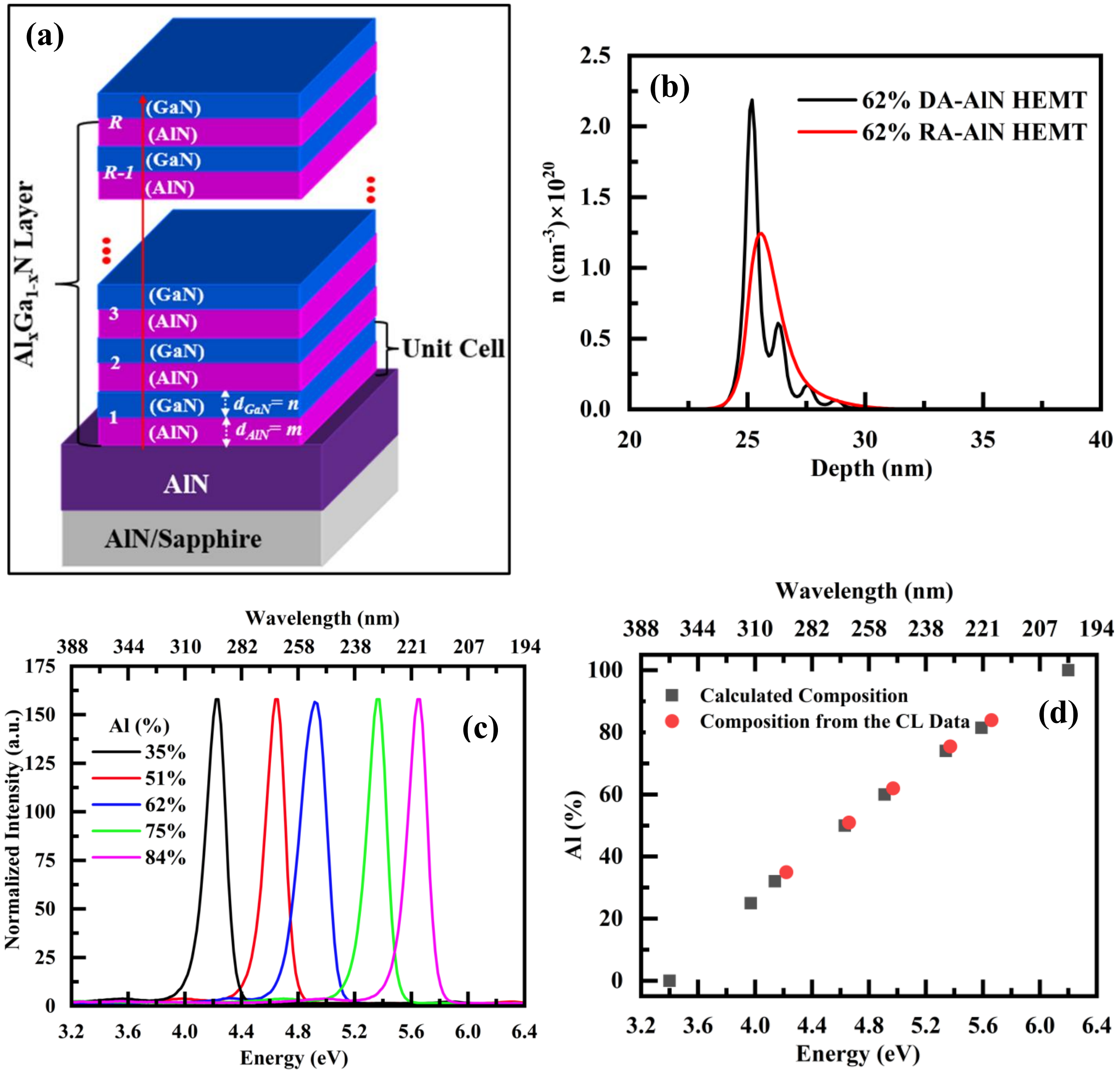

**Fig. 1** (a) AlN/GaN digital alloy unit cell, (b) Extracted carrier concentration for the random and digital alloy HEMTs using 1D Schrodinger-Poisson solver by G. Shnider[42] (c) RT Cathodoluminescence of the epi-layers at 20 keV, (d) Calculated and CL-estimated Al-alloy composition for the DA channels.

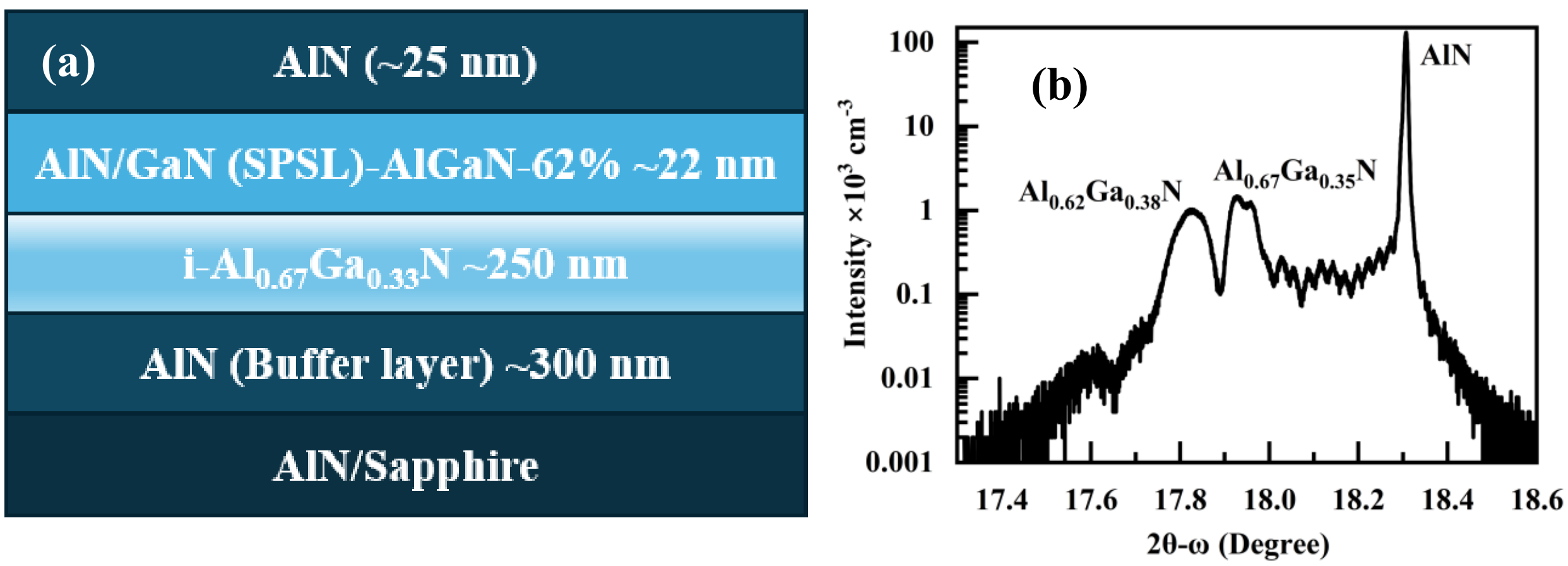

**Fig. 2** (a) Schematic of the epitaxial heterostructures, (b) The XRD omega-2theta (002) rocking curve for SPSL HEMT, which shows the intended peaks.

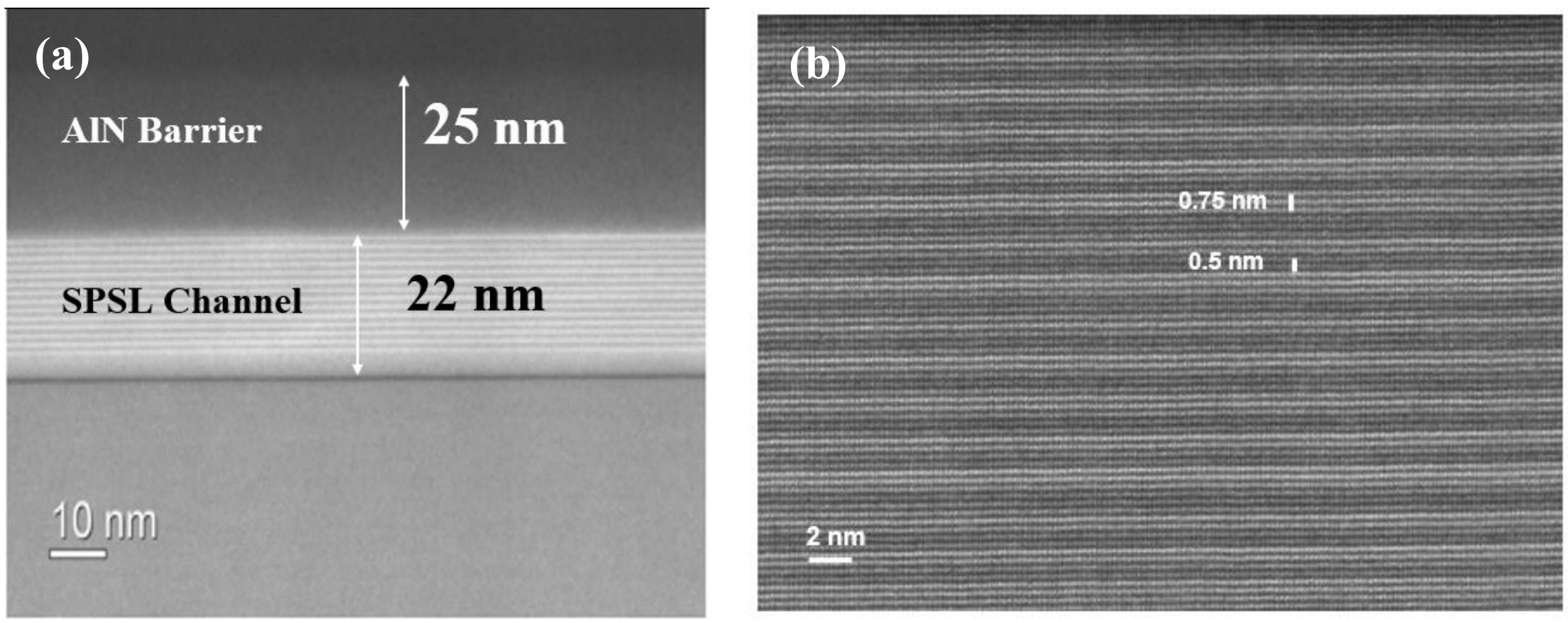

**Fig. 3** (a) High-resolution Bright Field (BF) cross-section TEM for the AlN/GaN SPSL and AlN barrier, (b) Zoom area of the SPSL AlN/GaN channel.

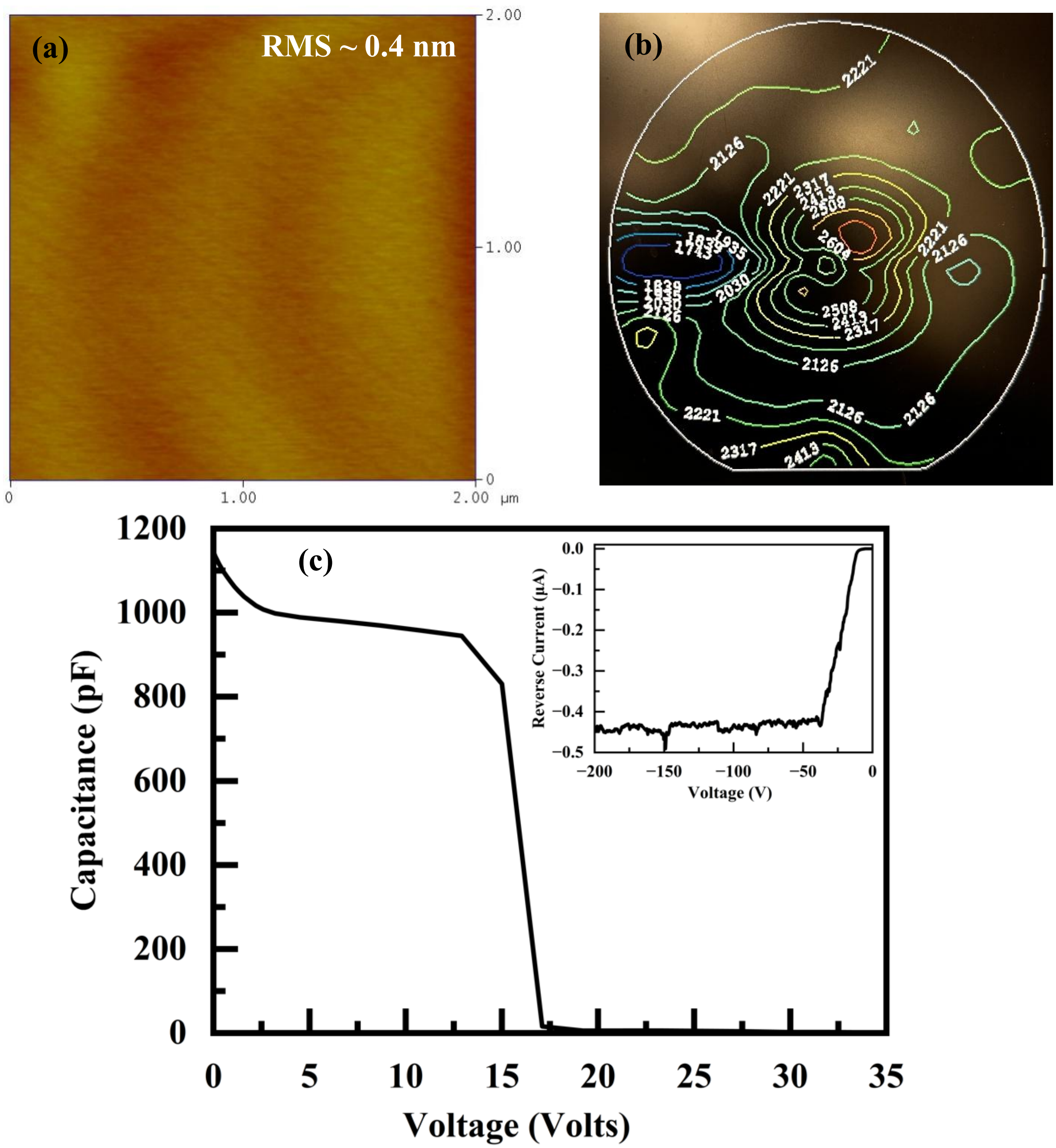

**Fig. 4** (a) Atomic force microscope scan of a 2 × 2 μm area of the epilayers, (b) Sheet resistance mapping for the 2-inch wafer, the average $R_{sh}$ ~2250 Ω/□ with 11% STD, (c) Mercury probe capacitance voltage (C-V) measurement data for the epi-layers at RT and 10 kHz frequency. Inset to Fig. shows the reverse current-voltage characteristic. As seen, the reverse bias current does not exceed 0.5 μA in the entire range of the applied voltages from 0 to 200 V.

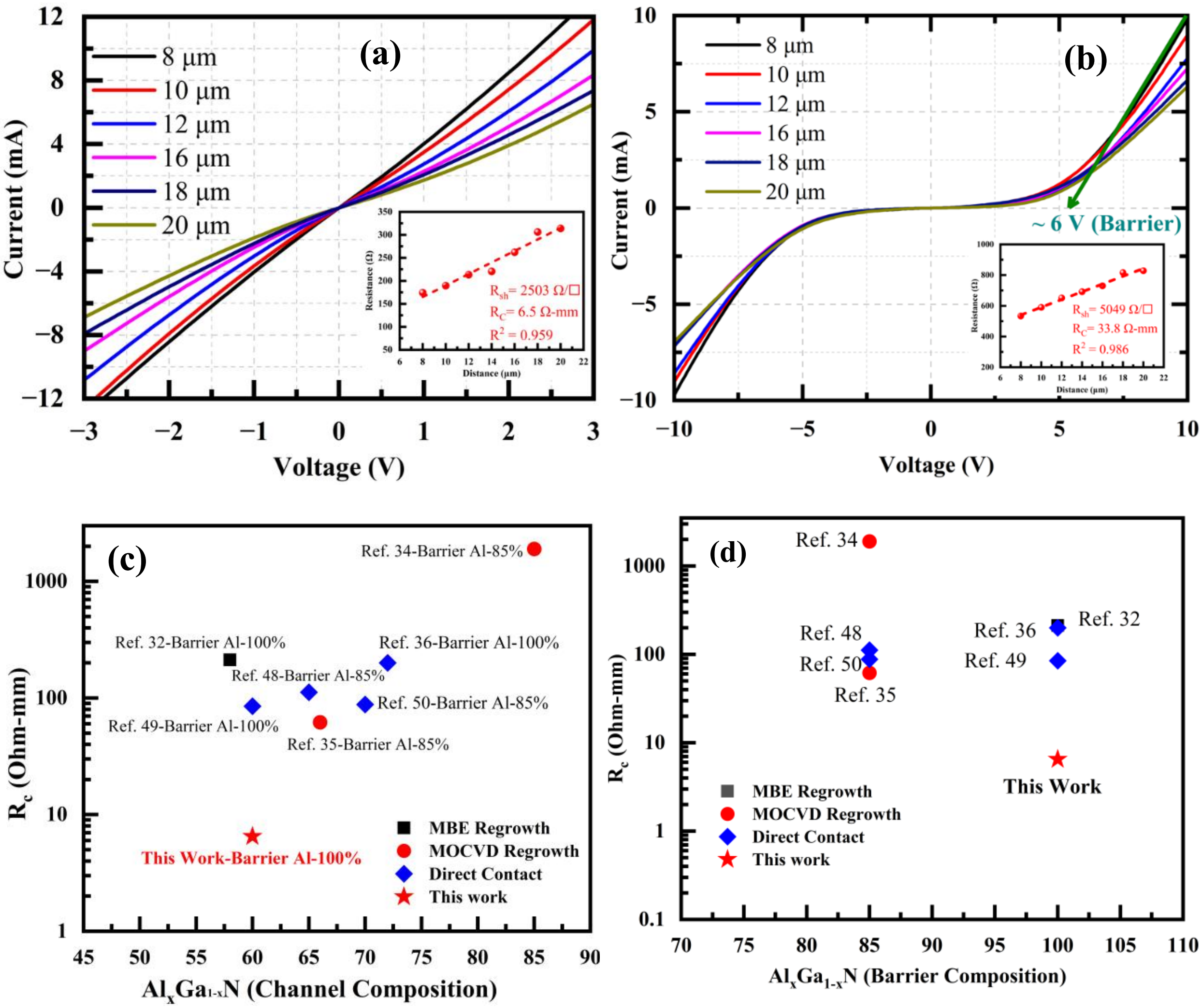

**Fig. 5** (a) RT I-V for the transmission length method (TLM) structures. Inset to the figure shows TLM electrode spacing versus resistance for the $R_c$ and $R_{sh}$ extraction. (b) RT I-V for the TLM structures with direct surface contacts. Inset to the figure shows TLM electrode spacing versus resistance for the $R_c$ and $R_{sh}$ extraction. (c) Benchmark comparison of the contact resistance ($R_c$) from this work with other Ultrawide Bandgap $Al_xGa_{1-x}N$ ($x > 0.5$) channel HEMTs with MBE or MOCVD regrown ohmic contacts. Plotted is the contact resistance versus the channel Al-alloy composition. (d) Contact resistance $R_C$ as a function of barrier alloy composition.[32, 34-36,48-52]